\documentclass[10pt,letterpaper]{article}
\usepackage[top=0.85in,left=2.75in,footskip=0.75in,marginparwidth=2in]{geometry}

\usepackage[utf8]{inputenc}

\usepackage{cite}

\usepackage{nameref,hyperref}

\usepackage[right]{lineno}

\usepackage{microtype}
\DisableLigatures[f]{encoding = *, family = * }

\raggedright
\usepackage{changepage}

\usepackage[aboveskip=1pt,labelfont=bf,labelsep=period,singlelinecheck=off]{caption}
\makeatletter
\renewcommand{\@biblabel}[1]{\quad#1.}
\makeatother

\usepackage{lastpage,fancyhdr,graphicx}
\usepackage{epstopdf}
\fancyheadoffset[L]{2.25in}
\fancyfootoffset[L]{2.25in}

\usepackage{color}

\definecolor{Gray}{gray}{.25}

\usepackage{graphicx}
\usepackage[ddmmyyyy]{datetime} 
\usepackage{sidecap}

\usepackage{wrapfig}
\usepackage[pscoord]{eso-pic}
\usepackage[fulladjust]{marginnote}
\reversemarginpar

\begin{document}
\vspace*{0.35in}
\captionsetup[figure]{font=footnotesize,labelfont=footnotesize}
% title goes here:
\begin{flushleft}
{\Large
\textbf\newline{Memory systems of the brain}
}
\newline

Alvaro Pastor 
\newline

Universitat Oberta de Catalunya, Computer Science, Multimedia and Telecommunications Department, Barcelona, Spain
\\
alvaropastor@uoc.edu
\bigskip
\\
January 1, 2020
\end{flushleft}

\section*{Abstract}
Humans have long been fascinated by how memories are formed, how they can be damaged or lost, or still seem vibrant after many years. Thus the search for the locus and organization of memory has had a long history, in which the notion that is is composed of distinct systems developed during the second half of the 20th century.
A fundamental dichotomy between conscious and unconscious memory processes was first drawn based on evidences from the study of amnesiac subjects and the systematic experimental work with animals. The use of behavioral and neural measures together with imaging techniques have progressively led researchers to agree in the existence of a variety of neural architectures that support multiple memory systems.
This article presents a historical lens with which to contextualize these idea on memory systems, and provides a current account for the multiple memory systems model.

% now start line numbers
\nolinenumbers

\section*{Introduction}

Memory is a quintessential human ability by means of which the nervous system can encode, store, and retrieve a variety of information \cite{Schacter2000MemorySO,Squire2009MemoryAB}. It affords the foundation for the adaptive nature of the organism, allowing to use previous experience and make predictions in order to solve the multitude of environmental situations produced by daily interaction. Not only memory allows to recognize familiarity and return to places, but also richly imagine potential future circumstances and assess the consequences of behavior. Therefore, present experience is inexorably interwoven with memories, and the meaning of people, things, and events in the present depends largely on previous experience.

Questions concerning the nature and organization of memory systems are fundamental to our understanding of the processes underlying learning and memory in humans and other animals. Due to its importance and complexity, the subject of memory is connected with some of the most profound questions of philosophy and biology, and as such, has been approached from different perspectives and methodologies. Along the way, idealists and materialists, empiricists and nativists, behaviorists and cognitivists, have confronted ideas about memory. First from philosophy through introspection and intuition, complemented later with anatomical and physiological observations. Then neuropsychology, through observation of patients with various defects in memory which developed the notion of a dichotomous organization of memory. More recently, relevant contributions from %cellular and molecular 
neurobiology and cognitive neuroscience, using various behavioral and neural measures together with imaging techniques, have progressively led researchers to postulate distinctions among various memory systems supported by different neural architectures, for example, by the hippocampus and related structures, the amygdala, the striatum, and the cerebellum.

The tradition of memory research considered in this article is that of experimental psychology and cognitive neuroscience amassed over the past decades. While a major part of early research assumed a unitary view on memory, some authors assumed dichotomous classifications based generally on the distinction of conscious and unconscious memory processes. Accordingly, relevant distinctions between memory systems have been proposed including but not limited to: episodic and semantic memory \cite{Tulving1972EpisodicAS,Tulving1983ElementsOE}, taxon and locale memory \cite{Okeefe1978TheHA},
habit memory and cognitive memory \cite{Mishkin1984MemoriesAH,Mishkin1984MemoriesAH2},
working memory and long term memory \cite{Baddeley1974WorkingM,BaddeleyBook86,Baddeley2000TheEB},
implicit and explicit memory \cite{Graf1985ImplicitAE,Schacter1987ImplicitMH},
declarative and non declarative memory \cite{Squire1996StructureAF},
declarative and procedural memory \cite{Cohen1993MemoryAA}, fast and slow memory systems \cite{McClelland1995WhyTA,Sherry1987TheEO}. 

In the last years, there is growing consensus about the nature and organization of memory. Supported on cognitive and neurobiological and neuropsychological evidence, memory is conceived as a set of separate, highly specialized, neural mechanisms that interact to reach a common goal. However, the complete extent of their features and interactions is yet to be fully understood. 

The first part of this article presents a brief historical review with which to reevaluate and contextualize the current theories and frameworks on human memory systems. The second part presents an account for the components of the multiple memory systems model, with some clues regarding the functional basis of memory processes. The memory systems comprised in this article are part of the most accepted and disseminated at present time. Debate about which of the different taxonomies of memory most appropriately organizes all available experimental data is beyond the scope of the present study.

\section{Historical background}

The nature of human memory and its organization have been object of study, speculation and debate for several centuries, in a variety of areas including philosophical inquiry, cognitive psychology, biology and neuroscience. In looking at the history of the localisation and organization of memory it is necessary to turn to ancient theories regarding the nature of soul and its relation to the body \cite{Engelhardt2020CerebralLO}. 

In early allusions to memory, ancient Egyptians considered it was solely supported by the heart, as the centralized seat of cognitive processes \cite{Ziskind2004ConceptsOT}. Contrary to other organs, the hearts of Egyptian mummies were kept carefully intact to ensure eternal life. Brains, on the other hand, were ideally removed via transnasal excerebration, a technique devised for this purpose by Egyptian embalmers \cite{TransnasalexcerebrationsurgeryinancientEgypt}.
In classical greek philosophy, memory was considered as part of the eternal and immaterial nature of the soul. In Theatetus, a major work written circa 369 BCE which  explores  the  nature  of  knowledge, Plato introduces the problem of remembering in the analogy of the wax into which our perceptions and thoughts stamp images of themselves. An early distinction was made between the passive retention of perceptions and active memory. Plato described active memory in its episodic features, as \emph{"the power which the soul has of recovering, when by itself, some feeling which she experienced when in company with the body"} \cite{BurnhamMemoryHA}. 
Later in the fourth century BCE, Aristotle returned to the notion that the soul is not a separate substance from the body, as Plato had taught, but inherently bound into it, affording its form. Aristotle came from a medical family and his biological bent is shown in its influence on most of his philosophy. Informed by his biological researches, in \emph{Metaphysica},  Aristotle presented a thoroughly materialistic theory, in which the soul or souls were distributed throughout the entire living body \cite{Aristotle1999AristotlesM}. Also in \emph{de Anima}, the soul was conceptualized as distributed in the body. While, according to Plato, memory is one of the higher faculties and partakes of the eternal nature of the soul, for Aristotle memory is dependent upon a physical process, and perishes with the body. 
In spite of this, Aristotle acknowledged the existence of a center on which sensory input converged and from which behavior was initiated. This, for Aristotle, is the location of common sense or \emph{sensus communis}, where besides the  external senses such as sight and smell were processed, internal senses such as memory and imagination were sustained. Aristotle established that the cognitive center was located at the heart, and that the vegetative soul was fluid and distributed everywhere in the body \cite{Bloch2007AristotleOM,draaisma_2000}.
The operation of memory as then conceived, may be inferred from the kinds of memory that the classical philosophers observed. Aristotle worked on a tract devoted to the subject of memory called \emph{De memoria et reminiscentia}. Its second chapter was devoted mainly to the recollection and association of ideas. In dichotomous fashion, Aristotle distinguishes the mere persistence and direct reproduction of a presentation, from voluntary recollection, which is possible only by the association of ideas, by similarity, contrast, and contiguity. In the Aristotelian view, persistence memory is an attribute of animals and humans, while the voluntary recollection is unique to humans \cite{Bloch2007AristotleOM,Presti2018LocalizingMA,aristotleonmem}.

Increasingly, philosophical inquiry on memory was influenced by anatomical and physiological studies. In the first century BCE the Byzantine surgeon Poseidonius produced the first written assignments of mental functions to specific cerebral regions. Based from observations of head injuries, Poseidonius argued that lesions at the front of the brain interfere with apprehension of all types of sensation, whereas trauma of the posterior part results in memory deficit, and damage to the middle ventricle produces a disturbance of reason \cite{poseid}. At the end of the fourth century CE, Nemesius of Emesa carried forward the ideas of ventricular function and proposed a schema which remained fairly standard for several centuries. In \emph{De Natura Hominis}, claimed that the anterior ventricles accounted for the mixing of sensations and for imagination, the middle ventricle for cogitation and reason, and the posterior ventricle for memory \cite{Parry2018LocatingMA,Panteleakos2013AnatomyAP}.
   
Augustine of Hippo, Roman theologian and philosopher at the end of fourth century CE, presented an account of human memory orchestrated by the brain and divided in different and complementary instances: sensible memory, intellectual memory, memory of memories, memory of feelings and passion, and memory of forgetting \cite{staugustine}.  Sense memory, for example, preserved and reproduced idealized images of visible objects, sounds, odors and other items available to the senses. Intellectual memory  on the other hand, contained our knowledge of the sciences, of literature and dialectic, and of the questions relating to these subjects.  This memory, unlike the memory of sense, contained not the images of things, but the things themselves \cite{Cassel2013FromAO}. The philosopher’s contribution foresees in striking detail some of the most relevant theories of modern psychology and neuroscience about memory, including the differences between sense memory and intellectual memory, similar to current episodic and semantic memory definitions. 

During the following decades, philosophical conceptions and directions in memory research in the Western tradition remained divided between either the views of Aristotle or those of Plato and Augustine of Hippo.

An important milestone is found in the mid 16th century. The work of Andreas Vesalius in \emph{De humani corporis fabrica libri septum}, helped developed an opposition to the medieval theory that the soul resides in the ventricles, believing these to be mere channels for the passage of the fluid \emph{anima} and explicitly identified the brain as the main organ of intelligence, movement, and sensation \cite{Vesalius2014TheFO}. Likewise, English physician Thomas Willis helped establish as well that activities are not carried out in the ventricles but in the brain substance itself. Also, Willis proposed a direct correlation between complexity of gyri convolutions and availability of cognitive abilities including memory. \cite{WraggeMorley2018ImaginingTS}. 

Willis' ideas were contemporary to those of the French mathematician and philosopher Descartes, and the notion of the correlation between physical and psychic processes was clearly understood by the dualists of the seventeenth century. In \emph{De homine} Descartes elaborated the mechanistic concept of human sensation and motion, and retained the ancient idea of the soul residing in the pineal gland, according to him located in the ventricles. \emph{ When the mind wills to recall anything,} Descartes wrote, \emph{this volition causes the pineal gland to incline itself successively this way and that, and impel the animal spirits to various parts of the brain until they come to that part in which are traces left by the object we would remember}. Descartes explained the physical processes involved in memory in accordance with his crude physiology, following the belief that the soul resides in the ventricles and that the fluid in them can be animated. This ideas prevailed into the end of the 18th century \cite{Urban2018OnMO}.

The doctrine of the Cartesians, that memory was conditioned by traces left in the brain, was developed by the physiologists of the 18th century into a materialistic and mechanical view of memory. In an extreme example, the physiologist Albrecht von Haller interested in the time occupied in psychic processes, estimated that a third of a second was sufficient time for the production of one idea. Von Haller and others then calculated that in a hundred years a human must collect during her wake period a total of 1 577 880 000 traces. By assuming the weight of the brain elements that were then conceived with the power for preserving impressions, then obtained 205 542 traces must be found in one gram of brain substance \cite{draaisma_2000}.

Early in 19th century, Franz Joseph Gall aimed again at settling the conceptual dispute, and contributed to establish that memory was located solely in the brain and was not located in the heart nor the ventricles as it was previously posited. Based on an anatomical and physiological data, Gall also acknowledged that the brain contains independent but interacting regions, and later interpreted that differences in behavior are due to differential development of these brain regions \cite{Buck2009TheDO}. To support his argument, Gall drew on observations of within and between individual differences in memory for particular kinds of information. Noting that some individuals have excellent memory for places but not music whereas others exhibit the opposite pattern, Gall contended such differences could not exist if memory constituted a unitary faculty \cite{Finger2019FranzJG,Eling2019FranzJG}. 
A few years later, physician Marie-Jean-Pierre Flourens investigated Gall’s controversial views on cerebral localization by removing anatomically defined areas of the brain of an animal and watching its behaviour. The results did not favour the idea of cerebral localization and led him to conclude that the brain functioned as a whole and thus arose the concept of \emph{cerebral equipotentiality} \cite{Pearce2009MarieJeanPierreF,frenolo}. Much debate ensued, due to the fact that his experiments were mainly on birds, and the techniques used were considered imperfect.

In contrast, French philosopher Maine de Biran published in 1804
a remarkable attempt to organize human memory within a taxonomic construct, distinguishing separate memory systems instead of a single unity: representative memory, that is the conscious recall of facts and events, mechanical memory, responsible for learning of habits and skills, and sensitive memory, regarding emotional information \cite{mainedebiran,Schacter1994WhatAT}. Psychologist and philosopher William James followed along these lines with his \emph{Principles of Psychology} in 1890. Despite no experimental evidences presented, a number of very important observations were made. The distinction between memory and habit, which led him to write separate chapters on both. Besides the distinction of habit, James also postulated two other different kinds of memory, establishing the difference between primary and secondary memory, which could be interpreted as an anticipation of the modern distinctions of short term memory (STM) and long term memory (LTM) \cite{Thompson1990TheNO,Hawkins2011WilliamJG}. 

Bergson’s reflections on memory in \emph{Matter and Memory} from 1896 also contributed to the modern non unitary view of memory. The French philosopher proposed distinctions between habit memory and episodic memories. For Bergson habit involved repetition, is effortless and, once formed, non-representational. Despite being non representational its effect percolates into the present and in part defines behavior. In contrast, what he understood as episodic memories were representational, involved a unique non-repeatable event, and its recollection, requires an act of will which involves an active letting go of the attachment to the present \cite{Bergson,Perri2014BergsonsPO}. In his 1949 work \emph{The Concept of Mind} the English philosopher Gilbert Ryle also distinguished between two types of memory: knowing-that, a type of memory that does not directly inform action but is expressed through regulative propositions; and knowing-how, a type of executive memory that cannot be built up from pieces of knowing-that and is actualized through enactment \cite{ryle}. These suggestive approaches to the nature and organization of memory and the subjective experience of recollection may be considered as the most influential in Western philosophy of the twentieth century.

While these are seen as the earliest attempts in the study of memory, what was demanded in order to fully understand its nature and organization was not only philosophical discourse or psychological intuition, but experimental inquiry.

In a major contribution in the study of memory, the psychologist Hermann Ebbinghaus published in 1885 \cite{Ebbinghaus2013MemoryAC} a pioneering experimental and quantitative treatment of memory, displaying new ideas and methods for rigorous and systematic controlled experiments on memory, that have had an enduring influence throughout the twentieth century of memory research.
A revolutionary concept in \emph{\"{U}ber das Ged\"{a}chtnis} was the idea that memory research  was not deemed to be the investigation of pre existing memories but that memory contents to be tested needed to be created and modified in the laboratory as an integral part of the experimental procedure \cite{ebbinghaus_2014,Ebbinghaus2013MemoryAC}. Ebbinghaus invented the paradigm of the nonsense syllable as learning materials of homogenous difficulty beyond the familiarity of words and prose. This notion defined a task in which a multitude of variables can be defined and their influences on remembering behaviors observed.
He used strict controls on the timing and number of study trials, allowed time for recall, and retention interval, and  measured the difficulty of learning a list by the number of study trials required to attain one correct recall. Moreover, Ebbinghaus  introduced the idea of measurable degrees of learning by noting the savings in relearning a list he had learned earlier. Using this measure Ebbinghaus was able to devise his famous forgetting curve relating percent savings to retention interval. \cite{Ebbinghaus1986HumanMA,Newman2012UpdatingEO}.

Even in Ebbinghaus' earliest experiments it was clear that learning depended greatly on the nature of the content, whether numbers, nonsense syllables, words, meaningful sentences or poetry, as these vary greatly in their familiarity and meaningfulness to subjects. By studying isolated variables, Ebbinghaus was able to observe the influence of the different types of learnt contents on memory performance, as well as the mode of their presentation, and the strategies subjects used in learning them. The memories established can then be tested by direct measurement of recall, recognition and reconstruction, or by a variety of indirect measures. Ebbinghaus' experimental framework set the foundations for the modern era of memory studies, and remains the prototype for its research almost 150 years after.

At the dawn of the twentieth century, two relevant contributions to the understanding of memory were presented in relation to the study of reflexes. The Russian neurologist and anatomist Vladimir Mikhailovich Bekhterev presented \emph{Pathways of brain and bone marrow} in 1900, exposing an association of memory with the hippocampus based on the study of amnesia in patients with hippocampal degeneration. In addition to his later \emph{Foundations for Brain Functions Theory} Bekhterev presented his views on the functions of the parts of the brain and the nervous system and developed a theory of conditioned reflexes which describe automatic responses to the environment called \emph{association reflex} \cite{Mishkin2016BehaviorismCA,Ertmer2018BehaviorismCC}.

Similarly, the scientific mainstream of behaviorism, with its main exponents Thorndike, Pavlov, Watson and Skinner, held  that all learning is the  attachment   of  responses  to  stimuli \cite{Clark2018LearningTB}. Therefore they studied the characteristics and components of a particular type of learning: that which is derived from repeated association between a stimulus and a response,named classical conditioning, and that which is derived from association between a stimulus and a behavior, called operative conditioning \cite{Mishkin2016BehaviorismCA,watson}.  Pavlov's   studies  are  related  to  a  type of  memory that later would be called associative memory, while Fitts and  Posner’s  studies \cite{Fitts1954TheIC}  are  considered  the  first aiming to explain procedural memory. It may be argued that the knowledge on memory produced by these experimental works of behaviorism was among the most important in the early days of twentieth century.

Guided by the then prevalent and straightforward view that stimulus-response learning is supported by connections between sensory areas in the posterior cortical areas and motor areas in the frontal cortex, Karl Lashley’s conducted pioneering efforts in the search for engrams \cite{Lashley1988InSO}. Engrams were viewed as fundamental to the understanding of memory, as the physical structures in which memory traces were stored. Using maze learning in rats he aimed to map the cortical areas and pathways that support memory formation. As the search proved unsuccessful, Lashley subscribed Flourens' view on the indivisible brain, meaning that there were no localizations of memory in different parts of the brain \cite {LashleyBrainMA}. He concluded that the memory trace was widely distributed both within cortical areas and throughout the cortex and that any of the neurons within cortical areas could support the engram, naming this principle \emph{equipotentiality}. Also contributed the notion that involved brain areas acted together to support the engram, naming this principle \emph{mass action} \cite{Orbach1982NeuropsychologyAL}.

Later in 1934 Donald Hebb, student of Lashley, followed the interest in how the brain gets organized to generate abilities such as memory. His 1949 work \emph{The Organization of Behavior} presented cell assemblies and the connection between cells that fire together, advancing studies of memory consolidation \cite{Shaw1986DonaldHT,Shepherd2009LearningAM}.  The \emph{dual-trace} hypothesis of memory subsequently proposed by Hebb in 1949 offered a specific physiological theory of memory consolidation.  According to Hebb’s hypothesis, neural activity initiated by an experience persists in the form of reverberating activity in neural circuits of STM, and that the formation of LTM means the stabilization of these circuits leading to permanent synaptic change \cite{McGaugh2000MemoryaCO}. 

During the following decades, as psychology struggled to assert itself as a scientific discipline, a large wealth of knowledge about memory  was supported on an increasing number of experimental studies. The work of Edward Tolman distinguishes from many of his behaviorist contemporaries on its focus on memory and behavior. Though Tolman was greatly influenced by Watson’s behaviorism, Tolman’s cognitive, anti-reductionist emphasis on purpose and macroscopic or molar behavior stood in stark contrast to many of the learning theories of his contemporaries \cite{TolmanMeansendreadinessAH,Tolman1935TheOA}.
The battle was sustained between these two schools of thought, stimulus-response learning versus cognitive learning, specially at the level of animal behavior. Tolman proposed that there is more than one kind of learning in animals and humans. Unlike motor patterns, well understood by behaviorists, other types of learning were postulated outside the stimuli-response logic. Importantly,  a distinction  between memory systems  is  rooted  in  this debate between place-learning, a position advocated by Tolman \cite{Tolman1948CognitiveMI}, and response-learning, a position advocated by Hull who asserted that learning was a general process \cite{Hull1943PrinciplesOB}. 

Such  behavioral findings  suggested  the  potential  co-existence  of  at  least two  distinct  learning  systems. However they did not have the effect of shifting the debate concerning how many memory systems there were. Other authors argued that these two kinds of learning existed, but that they did not represent fundamentally different things, instead they are accountable by the use of different cues \cite{restle57}.

A very important contribution of Tolman to the study of memory as set of dedicated systems was the \emph{Cognitive Map} theory, which provided deep insights into how animals represent information about the world and how these representations inform behavior.
Tolman developed this idea of a cognitive map as an alternative to the then-common metaphor of a central switchboard for learning and memory, typical of stimulus-response formulations. According to this view, animals maintained a set of complex, integrated expectancies or hypotheses of their world \cite{tolman32}. Tolman would later call these complex integrated expectancies cognitive maps \cite{Tolman1948CognitiveMI}. For Tolman, learning is acquiring expectancies, that is probability characteristics of the environment. The organism thus typically learns a cognitive map not a mere movement pattern. His later writings emphasized the use of multiple expectancies to inform behavioral performance, particularly when an animal is faced with a choice \cite{tolman_1966,carroll_2017}.

Consistent with the emerging understanding that processing and memory storage could not easily be separated from one another neurobiologically, the theory of the cognitive map was embodied within a multiple memory systems perspective. Importantly, it claimed that the brain memory systems differed based on the nature of the contents of the memories they processed, and not only based the length of time that memory storage was supported. As such, this view critically differed from the more traditional perspectives of cognitive psychology and cognitive neuropsychology prevalent a the time which saw systems as differing primarily in terms of how long information remained within them, not in terms of the type of information involved.

It is relevant for the study of memory systems that two kind of differences that render memory systems distinct were settled. First, the distinct memory systems might be computationally different at their circuitry level in order to be efficient, a notion later developed by Sherry and Shacter's theory of functional incompatibility \cite{Sherry1987TheEO}. Fodor also addresses the notion of modularity \cite{Fodor1983ModularityOM}, understood as separate domain specific modules that handle certain sort of contents. The functionalities of these modules are hardwired and not acquired through learning processes. Second, systems might differ in terms of the length of time during which information is to be stored within them, which can require different underlying mechanisms. Beyond the study of animals, these two non mutually exclusive approaches to memory systems were further expanded to humans in the study of focal cerebral damage.

Since the 1950's Scoville, Penfield, Hebb and Milner, documented the effects of surgical lesions of temporal lobe on human declarative memory, finding severe anterograde amnesia and retrograde amnesia with temporal gradient \cite{Scoville1954TheLL,Milner1955TheEO,Penfield1958MemoryDP,Scoville1957LOSSOR,Friedman1969ChronicEO,Hebb1940HUMANBA}. A well known case of study was that of the severe epileptic patient Henri Molaison (H.M.) who in 1953 underwent surgical removal of brain structures in the MTL (MTL) including the hippocampus to cure him of his epileptic seizures. As a result of this surgery, H.M. was profoundly impaired in the acquisition of certain forms of information. He became unable to consciously recollect new events in his life or new facts about the world. However, H.M. retained the ability to remember other types of information. While other cognitive functions remained intact, H.M.'s experience was greatly influenced by a permanent present tense. If rehearsed and attention was held, a patient such as H.M. could keep a short list of numbers in mind for several minutes. But would prove unable if distracted or when the amount of material to be remembered exceeded what can be held in immediate memory.

Early evidence from H.M's case supported the mainstream view that memory is composed of systems which differed in terms of how long can they keep the memory for. By having removed large portions of hippocampus H.M. had only distorted the consolidation of the LTM system \cite{Milner1958PsychologicalDP}.% the ability to transport memory materials from immediate memory to intact short-term memory

Subsequently, Milner reported that H.M. could learn and perform new sensorimotor tasks.  Although H.M.'s memory impairment disabled learning for all forms of materials, including scenes, words or faces, he proved capable of learning a complex hand – eye coordination skill, namely mirror drawing, over a period of 3 days even without awareness of having been previously trained or tested \cite{Milner1962LesTD}. This selectivity of H.M.'s amnesia related to the type of learned activity or content, suggested the existence of two distinct forms of memory corresponding to each type of material: declarative memory and procedural memory. It was presumed that these systems were dependant on separate neural systems, and that by lesion or degenerative processes, amnesiac patients were only disabled in their declarative memory functions \cite{Milner1963EffectsOD,Milner1966AmnesiaFO,Baxendale1998AmnesiaIT}. 

These findings contrasted with prevailing views at the time by linking a memory disturbance to damage to a specific part of the brain. However these theories were attempted in animal models, namely monkeys and rats, leading to no amnesia, thus discussions and uncertainties about a content based organization of memory systems ensued for several years \cite{Freeman1939AnIO, Finan1939EFFECTSOF,Orbach1960LearningAR,kimble1953psychology}.

 In the following sections an account is provided regarding these two kinds of taxonomies and a brief review of the relations and differences between memory systems.
 
\section{Persistence-based taxonomy}

The 60's saw a growing interest in developing mathematical models of learning and memory. Despite that the mainstream view accepted only one global memory system responsible for handling many kinds of information, this assumption did allow an implicit division of memory in memory systems which differed in terms of their capacity to store contents within them.

In the previous section, a review on the historical evolution showed that allusions to memory taxonomies based on time persistence had been present for several centuries, despite the apparent triviality of its insight on how memory is organized. One possible catalyst for the emergent importance of this perspective during the early 60's, was the influence of the computer revolution which may have served as inspiration for thinking about memory from an information processing approach.

In its most influential consolidation, Atkinson and Shiffrin \cite{Atkinson1968HumanMA} settled in 1968 a model of human memory divided into three classes, ultra short-term memory, STM, and LTM, which became known as the multi store or modal model of memory. %\ref{figprocessing}.
In addition to identifying the three different memory types, Atkinson and Shiffrin also suggested that they stand in a rather fixed relationship to one another. They proposed that stored information is processed sequentially entering the brain through the sensory system into ultra-short-term, passes through STM and ends up in LTM. This processing chain can terminate when information is forgotten from one of the memory stores.
\begin{figure}[ht] 
\includegraphics[width=\textwidth]{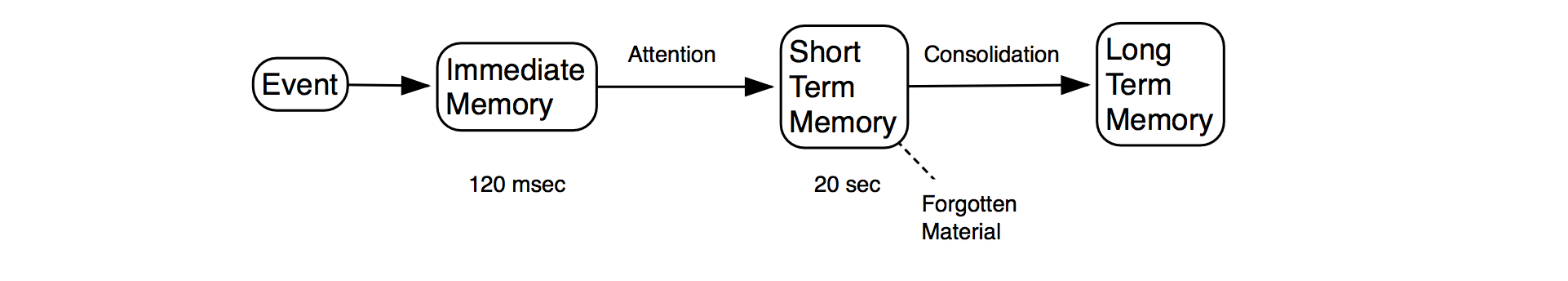}
\caption{\color{Gray} \textbf{ In the 1950's and early 60's the conception of memory was largely inspired by an information processing approach, in which systems differed in terms of duration, processing and storing information for varying periods of time.}
}
\label{figprocessing} 
\end{figure}

\subsection{Ultra short term memory}
According to this model, Ultra short term memory, or sensory memory, makes instantaneous sensory information available for processing for less than 120 milliseconds \cite{Coltheart1980}. This kind of memory is specific to a particular sensory system and does not appear to be shared across sensory systems. In the visual system, ultra short term memory is the reason, why, for example, apparent continuous motion is perceived when watching a movie instead of 24 individual frames per second. 

Authors have suggested that ultra short term memory is limited to only a few hundred milliseconds in duration because the memory traces decay naturally \cite{Atkinson1968HumanMA}. Due to this timescale this form of memory would operate outside of awareness, and since it is entirely driven by external stimuli, it lies outside conscious control. For these reasons, some researchers suggest that ultra short term memory may be a component of the sensory system rather than a memory system.

\subsection{Short term and Working memory}
In the late 50's the consolidation of the notion of a STM system began with the empirical support which observed that, provided the prevention of active rehearsal, even small amounts of information showed rapid forgetting \cite{Brown1958SomeTO,Ricker2016DecayTO}. The pattern of memory decay appeared to differ from that observed in standard LTM experiments, which led to the view that performance depended on a separate short term store. It was then thought that STM operated as a crucial antechamber of LTM storage. 

During the 60s, methods were developed for studying visual perception and motor responses with single-cell recordings from awake, behaving monkeys 
performing a classic delayed response task, which requires holding information in memory for a brief period. This work identified cells in prefrontal cortex that were maximally active during the delay portion of the task of 15 to 60 seconds \cite{Fuster1971NeuronAR,Atkinson1968HumanMA,Fuster2008AnatomyOT,Williams1995ModulationOM}.

However, various problems stemmed from the view of STM held by this model. First, it was challenged the notion that the probability of an item being stored in LTM as a function of its maintenance on short term system. A number of studies demonstrated that active verbal rehearsal was not linked to durable LTM \cite{Craik1973TheRO}. This argument gave way to the theory called \emph{levels of memory} \cite{Craik1972LevelsOP}. This theory proposed that an item in order to be remembered, for instance a word on a page, could be processed at a series of encoding levels which started with the visual appearance of the item, its sound when pronounced, and later its meaning and relationships with other related memory experiences. A second challenge was provided by the neuropsychological evidence of patients with an STM deficit while preserving LTM intact. The then prevalent STM model failed to provide explanation for this phenomena, as it argued that poor performance of STM should have also led to impaired LTM capacity. These challenges led researchers to reformulate the STM hypothesis and postulate a multicomponent system which was termed working memory.

Working memory refers to the capacity to maintain temporarily a limited amount of information in mind, which can then be used to support various abilities, including learning, reasoning, and preparation for action \cite{Baddeley1974WorkingM}. It is generally acknowledged that working memory can store information for up to 20 seconds, and according to some experimental studies its maximum capacity is four chunks of information \cite{Cowan2001TheMN}. Despite its acquisition being without a conscious effort, working or short-term memory can be controlled consciously and is less dependent on sensory inputs than ultra-short-term memory.  Most importantly, through rehearsal, the spontaneous decay of working memory may be prevented.

According to Baddeley and his tripartite model \cite{BaddeleyBook86,Baddeley1994DevelopmentsIT} working memory is comprised by the phonological loop, which support retrieval temporary storage and rehearse of phonological material, the visuospatial sketchpad to perform maintenance of visual representation of stimuli and their position in space\cite{Baddeley1974WorkingM,BaddeleyBook86}, and a central executive sub system which is a limited capacity supervisory attentional system \cite{Shallice1982SpecificIO} that orchestrates and mediates the manipulation of contents from visuospatial and phonological  sub systems.
However, there is evidence indicating that spatial and object  working memory may be separable \cite{Baddeley1994DevelopmentsIT}. Single cell recording studies in nonhuman primates have revealed different activations in ventral and dorsal prefontal cortex during object working memory performance for the first and spatial working memory performance for the latter.

In 2000 Baddeley included in this model the episodic buffer involved in linking information across domains to form integrated units of visual, spatial, and verbal information with episodic chronological ordering. This episodic buffer is also assumed to have links to LTM and semantic meaning. The working memory model currently consists of four elements that process information: the central executive for attention control; the visuospatial sketchpad, which creates and maintains a visuospatial representation; the phonological buffer, which stores and consolidates new words; and the episodic buffer which stores and integrates information from different sources \cite{Baddeley2000TheEB}.

%\ref{figptripartite} 
\begin{figure}[ht] 
\includegraphics[width=\textwidth]{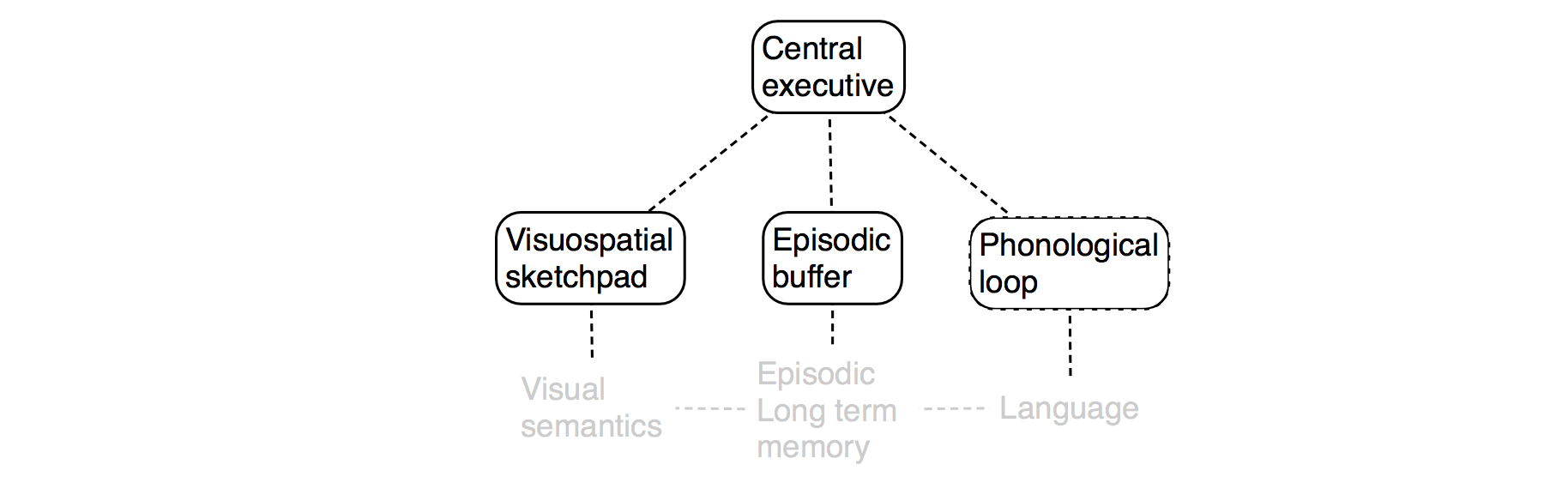}
\caption{\color{Gray} \textbf{ The year 2000 version of the multi-component working memory model according to Baddeley which includes the episodic buffer. The episodic buffer is interpreted to be controlled by the central executive, and it is capable of multi-dimensional information storage, providing a temporary interface between the phonological loop, the visuospatial sketchpad and long term memory (LTM). The buffer serves as a modelling space that is separate from LTM, but which forms an important stage in long- term episodic learning. In this schematic view, faded areas represent systems capable of accumulating long-term knowledge, and unshaded areas represent fluid capacities such as attention and temporary storage.
}
}
\label{figtripartite} 
\end{figure}

Working memory is therefore central to the ability to select and implement goal-directed behavior, to exercise what are termed executive functions. Indeed, recent discussions emphasize a broad role of prefrontal cortex in cognitive control, an idea that implies top-down influences from prefrontal cortex that direct attention and organize action \cite{Fuster2001ThePC,Rich2009RatPC}. The prefrontal cortex allows memory to be accessed strategically, and it orchestrates the use of learned rules so that knowledge relevant to current goals can be brought to mind and put to flexible use.

\subsection{Long term memory}
Assumed as a group of systems responsible for the capacity to store information over long periods of time \cite{Squire1992DeclarativeAN}.
LTM differs from STM in that the memory contents would no longer be vulnerable to spontaneous decay, nonetheless, some memories may be forgotten due to interference from memories that are stored later. Another key difference to short-term memory is that subjects recall these items not in isolation, but including the relationships between them, and the spatiotemporal context in which they are embedded.  This makes LTM singular due to its larger capacity for the storage of arbitrary combinations of different kinds of information \cite{Baddeley2013EssentialsOH}. 

Analysis indicates that the hippocampus and associated structures of the MTL serve to generate LTM traces. Clinical evidence indicates that damage to the hippocampus produces anterograde amnesia. At the cellular-molecular level, the intricacies of synaptic plasticity, a candidate model for memory storage, are being studied in great detail \cite{squire_kandel_1999}, however, the link between single neuron computation and computation at the network level is poorly understood. It remains unclear how neuronal cooperativity in intact networks relates to memories or how network activity in the behaving animal brings about synaptic modification\cite{Buzski2002ThetaOI}.

In recent years, psychologists have gathered numerous pieces of evidence for the view that our memories are in fact constructions of what happened in the past. Memories are stored in the brain in a distributed pattern in the outer layer of the cortex, related to the area of the brain that initially processed them \cite{Squire1992DeclarativeAN}. In this view, a visual aspect of an experience is normally stored in the visual cortex, an auditory aspect of an experience is stored in the auditory cortex, and a motion element of an experience is stored in the sensory motor cortex. When a memory is retrieved, all its fragments are put back together rarely identical to the initial experience \cite{Eichenbaum2001FromCT}.

In past decades, little thought was given to the possibility that there might be some constraint on the kind of information these memory systems could handle or the rules by which they operate. Over the years, several aspects of Atkinson and Shiffrin’s taxonomy have been questioned. The main criticisms can be divided into two classes. First, their taxonomy artificially splits memories of the same kind into different taxa. Several authors have proposed that STM and LTM are instances of the same unitary memory that happen to have different properties. Second, Atkinson and Shiffrin associate together disparate kinds of memories into a single system. Furthermore, The dominant view of a single memory system prevented the selective nature of amnesia from being systematically explored for several decades. For the most part neuropsychologists working with human subjects accepted the single system premise and the view that damage to the brain to varying degrees accounted for the global amnesia observed in subjects with various etiologies \cite{Warrington1968ASO,Warrington1968NewMO}. Despite these controversies, Atkinson and Shiffrin’s taxonomy has been one of the most influential taxonomies of memory.

Advances in neurobiology and behavioral neuroscience contributed to modify this approach. During the 70s a genuine change occurred in the way  memory is thought of, and various new phenomena began to be addressed as memory related, using different tasks to measure different specialized functionalities. Perceptual memory or representation systems are an example of this broadening \cite{Raskin1971LongTermPM}. Phenomena like object priming or fragment completion, added to the view that processing and memory storage occur in the same circuits has emerged from the neurobiological literature. \cite{Schacter1977MemoryFA,Tulving1982PrimingEI,Tulving1988PrimingOS}. Importantly, the discovery of the declarative memory systems with the view that the hippocampus has a selective role in learning, began in the late 60s and took more than 20 years to unfold.

\section{The discovery of declarative memory systems}
The first observations of H.M., and the results of formal testing, were reported in 1957 by Scoville and Milner \cite{Scoville1957LOSSOR} in a seminal paper that shed light on three fundamental principles: a) memory is a distinct brain function, separable from general cognitive and language functions; b) the MTL is not necessary for immediate memory since this was largely normal in H.M;  and c) MTL structures are not the final storage location for memories, since although the removal of these structures showed severe amnesia in H.M., the patient still retained large portions of memories of his childhood \cite{Squire2011TheCN}. This publication became one of the most cited papers in neuroscience and is still frequently cited. 

Importantly, subsequent findings from H.M. established that memory was not a monolithic system that handles all types of information, but a highly interconnected set of systems. Due to the the fact that H.M. and other amnesic patients demonstrated intact learning and retention of certain motor, perceptual and cognitive skills, memory was thought to be comprised of at least two separate systems and that only one kind of memory, declarative memory, is impaired in amnesia \cite{Cohen1980PreservedLA, Milner1962LesTD, Milner1966AmnesiaFO, Milner1968FurtherAO}.

Nevertheless, patients with hippocampal damage typically have a normal vocabulary, and their general knowledge of facts remains intact. Interestingly, the opposite pattern of memory loss is associated with a specific form of neurodegenerative disease known as \emph{semantic dementia} in which recent events are retrieved accurately while the meaning of words and knowledge about world facts is greatly impaired \cite{Patterson2007WhereDY}.

\subsection{Declarative memory}
Declarative memory provides a fast and flexible way to represent the external world, and the ability to make inferences from and generalizations across facts derived from multiple processing sources. It supports the encoding of memories in terms of relationships among multiple items and events. Declarative memory allows remembered material to be compared and contrasted. The stored representations are flexible  and can guide performance under a wide range of test conditions \cite{Squire2004MemorySO}. A key characteristic is that declarative memory is explicit and accessible to conscious awareness. That is, it may be brought to mind verbally as a proposition, or non verbally as an image 
\cite{Squire1991TheMT,Squire1992DeclarativeAN,Squire1996StructureAF,Baddeley2000TheBD}. 

It is widely accepted that declarative memory has two major components: semantic memory, responsible for processing facts about the world; and episodic memory, the ability to re-experience a time and place specific event in its original context \cite{Tulving1983ElementsOE,Schacter1994WhatAT}.

Declarative memory depends on the integrity of neural systems damaged in amnesia, namely the hippocampus and MTL associated structures. Amnesic patients' defects in memory extend to both verbal and nonverbal material, and involve information acquired through all sensory modalities. Other etiologies of amnesia have also contributed useful data to support theories on declarative memory, including encephalitis, anoxia and ischemia, Korsakoff's syndrome, and psychiatric patients
whose memories were impaired as a result of electroconvulsive therapy
\cite{Gabrieli1998ImagesOM,Preston2005HippocampalFD}.

Several theories regarding the role of the hippocampus in memory have been proposed over the years. All regard the hippocampus as being critical for episodic memory, but there are key differences in whether they view the hippocampus as having a time limited role in episodic memory and in whether they deem it to be necessary for the acquisition of non-contextual information. The pattern of spared and impaired cognitive processes in patients with hippocampal damage, combined with results from animal models of amnesia, has lead to the \emph{Declarative Theory} of hippocampal function \cite{Squire1969MechanismsOM}. Some researchers have focused on building a neural level understanding of hippocampal function in a specific cognitive domain such as the theory of the \emph{Cognitive Map} \cite{Okeefe1978TheHA}. While others have sought more specific characterizations of hippocampal function, drawing on experimental data from animals and humans with the \emph{Multiple Trace Theory} \cite{Nadel1997MemoryCR}, the \emph{Dual Process Theory} \cite{Aggleton1999EpisodicMA}, and the \emph{Relational Theory} \cite{Cohen1997MemoryFI}.

In addition to MTL structures the acquisition of episodic memory requires the involvement of other brain systems, especially the frontal lobes \cite{Tulving1989MemoryPK,Shimamura1991WhatIT,Halsz2016TheRO,Stebbins2002AgingEO,Moscovitch2005HippocampalCC}.

A number of considerations suggest that the capacity for declarative knowledge is phylogenetically recent, reaching its greatest development in mammals with the full elaboration of medial temporal structures, especially the hippocampal formation and associated cortical areas. This capacity allows an animal to record and access the particular encounters that led to behavioral change. The stored memory is flexible and accessible to all modalities \cite{Tommasi2009CognitiveBE}.

The declarative memory system has a distinct developmental evolution through individual's lifetime. In most species hippocampus is not fully developed at birth and processess of neurogenesis are constant throughout life in some cases \cite{Praag2002FunctionalNI}. In humans the hippocampus and related structures incur in a dramatic size change in the first two years of life, enabling the memory function after 18 to 24 months, and only reaching full adult function after reaching 10 to 12 years of age. As such, phenomena such as infantile amnesia defined as the inability to form episodic memory early in life is starting to be understood \cite{Nadel1984InfantileAA,Travaglia2016InfantileAR,Madsen2016OntogenyOM}. Similarly, in spatial navigation tasks, children younger than two years old fail at distal cue use in navigation, as it requires a fully functioning hippocampus \cite{Tommasi2009CognitiveBE,Edgin2014RememberingTW,Edgin2019TheH}.

\subsection{Declarative systems}

Tulving proposed the distinction between Episodic and semantic memory as two types of declarative memory subsystems. Both types of memory are declarative due to the fact that retrieval of episodic and semantic information may be carried out explicitly and subjects are aware that stored information is being accessed \cite{Tulving1972EpisodicAS, Tulving1983ElementsOE, Tulving1991ConceptsOH}. This distinction devised by Tulving has in many ways fulfilled its purpose of understanding, and accounting for, the broader range of memory phenomena and experimental findings, and has formed the foundation for decades of theoretical and experimental work in the cognitive neuroscience of memory.

 Of particular interest has been the extent to which semantic and episodic memory have a shared dependence on the hippocampus. In contrast to the definitive evidence for the link between hippocampus and episodic memory, the role of the hippocampus in semantic memory has been a topic of considerable debate. Amnesic patients do have great difficulty acquiring semantic knowledge, but they can typically succeed to some extent after much repetition \cite{Glisky1986ComputerLB,Bayley2008NewSL}. In Tulving's work on the severely amnesic patient K.C., he reports how eventually the patient learned to complete arbitrary three-word sentences during a large number of training trials distributed over many months despite the absence of any memory at all for training episodes \cite{Tulving1991LonglastingPP}. Episodic memory can be virtually absent in some severely amnesic patients who can still accomplish some semantic learning.

The key structures that support declarative memory are the hippocampus and the MTL adjacent structures, such as entorhinal, perirhinal, and parahippocampal cortices, which make up much of the parahippocampal gyrus \cite{Squire1991TheMT}. These structures are organized hierarchically, and their anatomy suggests how the structures might contribute differently to the formation of declarative memory, for example, in the encoding of objects in perirhinal cortex, or scenes in parahippocampal cortex, and in the forming of associations between them in the hippocampus
\cite{Squire2004TheMT,Staresina2011PerirhinalAP,Davachi2006ItemCA
}.

 One interpretation is that both episodic and semantic memory depend on the brain system damaged in amnesia, the hippocampus and MTL related structures, and that episodic memory additionally depends on the integrity of the frontal lobes. Patients with frontal lobe damage, who are not amnesic, exhibit a phenomenon termed \emph{source amnesia} which refers to loss of information about when and where a remembered item was acquired \cite{Janowsky1989SourceMI,Shimamura1987ANS}. In this sense, source amnesia appears to reflect a loss of episodic memory related to frontal lobe dysfunction, which in turn reflects a disconnection between facts and their contexts. The greater contribution that frontal lobe function makes to episodic memory, compared to semantic memory, would give a 1biological support to the distinction between episodic memory and semantic memory \cite{Squire2004MemorySO}.

In their infuential book \emph{The Hippocampus as a Cognitive Map}, O’Keefe and Nadel suggested that the then recently discovered place cells found within the hippocampus provided the neurobiological substrate of the \emph{Cognitive Map} \cite{Okeefe1978TheHA}. Subsequent investigations of hippocampal place cell activity further developed the notion of the hippocampus mediating place information and the functions of such data coordination beyond spatial domains \cite{redish1999}.

\subsubsection{Episodic memory}

Episodic memory was conceived by Tulving \cite{Tulving1972EpisodicAS}, at a time when information processing models dominated. Episodic memory refers to autobiographical memory for events in its when, what and how components, linked together in a coherent spatial and temporal context. This includes the spatiotemporal relations between events. A key feature of episodic memory is its unique role to allow the individual to mentally travel back into her personal past \cite{Suddendorf1997MentalTT,Eichenbaum1994TwoFC}.

 The precise details that characterize episodic memory have been, and continue to be, debated, but time, space and sense of self are widely accepted as key elements \cite{Tulving2002EpisodicMF,Schacter2012TheFO,Howard2015TimeAS,Schiller2015MemoryAS, Hassabis2007PatientsWH}. 
 
Episodic memory is affected by aging processes \cite{Lundervold2014AgeAS}, and  there have been findings of sex differences in its performance \cite{Pauls2013GenderDI,Herlitz1999SexDI}, but results are still inconclusive \cite{Persson2015AgeAS}.

It  has  been  suggested  that  episodic  memory  in  the hippocampus  is  formed  by  combining  spatial  information  from  the  medial  entorhinal cortex \cite{Hafting2005MicrostructureOA,Savelli2008InfluenceOB}  with non spatial information  from  the  lateral  entorhinal  cortex \cite{Hasselmo2009AMO,Hayman2008HowHP}. A number of imaging studies have revealed evidence linking MTL activation with episodic encoding. MTL activation has been observed under conditions in which exposure to novel stimulus materials is compared with exposure to familiar materials. Imaging studies have also illuminated the contributions of distinct prefrontal regions to encoding and retrieval \cite{Shallice1994BrainRA,Frith1997BrainMA,Kirwan2008ActivityIT}.
 
Recent studies have helped to establish that these neural systems  also support the capacity to imagine and to simulate episodes expected to occurr in the future \cite{Tulving1983ElementsOE,Suddendorf1997MentalTT, Schacter2007RememberingTP,Maguire2011RoleOT}.
The ability to pre experience future events has been referred by some authors as prospection \cite{Buckner2007ProspectionAT} and by others as episodic future thinking \cite{Atance2001EpisodicFT,Atance2008FutureTI,Atance2019ThinkingAT}.

Some early observations along these lines were reported related to the patient K.C., who was investigated extensively over 20 years since a head injury left him with large bilateral hippocampal lesions which caused a remarkable case of memory impairment \cite{Tulving1988PrimingOS,Tulving1991LonglastingPP,Rosenbaum2005TheCO}. K. C. was unable to provide a description of his personal future for any time period, immediate or distant, describing his mental state as \emph{blank}. Interestingly he used exactly the same definition of mental state when asked to think about the past \cite{Tulving1985HowMM,Tulving1988PrimingOS}.
A later investigation in another patient, D. B., who became amnesic as a result of cardiac arrest and consequent anoxia revealed that he, like K. C., exhibited deficits in both retrieving past events and imagining future events \cite{Klein2002MemoryAT}.

A study led by Hassabis examined the ability of five patients with documented bilateral hippocampal amnesia to imagine new experiences, such as being at a desired location in an ideal situation \cite{Hassabis2007PatientsWH}. 
Based on the content, spatial coherence and subjective qualities of the participants’ imagined scenarios, the ability to imagine of the subjects with hippocampal lesions was greatly reduced in richness and content.

Due to the fact that hippocampal amnesics have difficulty imagining new experiences, Hassabis concluded that the hippocampus plays a key role in recombining details of previous experiences into a coherent new imagined construction.

This interpretation that hippocampus mediates both future episode thinking and imagination, assumes that the episodic memory is constructive rather than a reproductive system. 
Since the future is not an exact repetition of the past, it requires a system that can draw on the past to flexibly extract and recombine elements of previous experiences. In this view, fundamental features of a memory are distributed widely across different parts of the brain \cite{Squire2004TheMT}. Therefore, retrieval of a past experience involves a process of pattern completion, in which the process of recall pieces together some subset of distributed features that comprise a particular past experience \cite{Marr1971SimpleMA,Norman2003ModelingHA}.

Consistent with this approach, neuroimaging studies focused on the brain activity during the construction of past and future events, have revealed regions exhibiting common activity, which included the left portion of the hippocampus and the right occipital gyrus, especially during the elaboration phase, when participants are focused on generating details about the remembered or imagined event \cite{Tsukiura2002MedialTL,Szpunar2007NeuralSO}.

\subsubsection{Semantic memory}
Tulving postulated that semantic memory consists of a \emph{mental thesaurus} encompassing a wide range of organized information including facts, concepts and vocabulary necessary for language \cite{Tulving1983ElementsOE,Tulving1991ConceptsOH}. Semantic memory can be distinguished from episodic memory by virtue of its lack of association with spatiotemporal contexts \cite{ Schacter2000MemorySO}, and does not require subjective reexperiencing of the episode in which the knowledge was acquired. These systems differed as well in terms of the conditions and consequences of retrieval. Unlike episodic memory, retrieval from semantic memory was thought to leave its contents unaltered and to provide new input into episodic memory. Retrieval from episodic memory, in contrast, was thought to modify the contents of the system. Also, episodic and semantic memory were thought to differ in their dependence on each other: whereas semantic memory was thought to be independent from episodic memory in terms of recording and maintaining information, episodic memory could be strongly influenced by, and depend on, information in semantic memory at encoding \cite{Greenberg2010InterdependenceOE}.

Originally, Tulving viewed episodic and semantic memory as two functionally separate systems, but he also emphasized the features shared between them and their close interaction. For example, retention of information in both systems was thought to be automatic, because it did not require ongoing effort; for both systems, retrieval of information could be prompted by highly relevant cues or questions, even if the retrieval process itself is outside of awareness. Personal semantic memory, that is semantic knowledge that relates to the self, represents another common area of semantic and  episodic memory, increasingly supported by neuropsychological evidence  \cite{Renoult2012PersonalSA,Grilli2014PersonalSM}

 Semantic memory has been associated with an overlapping network of neural regions, with recent research invested in elucidating the roles of anterior and lateral temporal neocortex and ventrolateral pre frontal cortex \cite{ThompsonSchill1997RoleOL,AtirSharon2015DecodingTF,Coutanche2015CreatingCF}. Imaging studies have suggested that the semantic attributes of a stimulus are stored near the cortical regions that underlie perception of those attributes, while have also highlighted that retrieval from semantic memory and encoding into episodic memory share underlying component processes \cite{Cappa2008ImagingSO}.

 \begin{figure}[ht] 
\includegraphics[width=\textwidth]{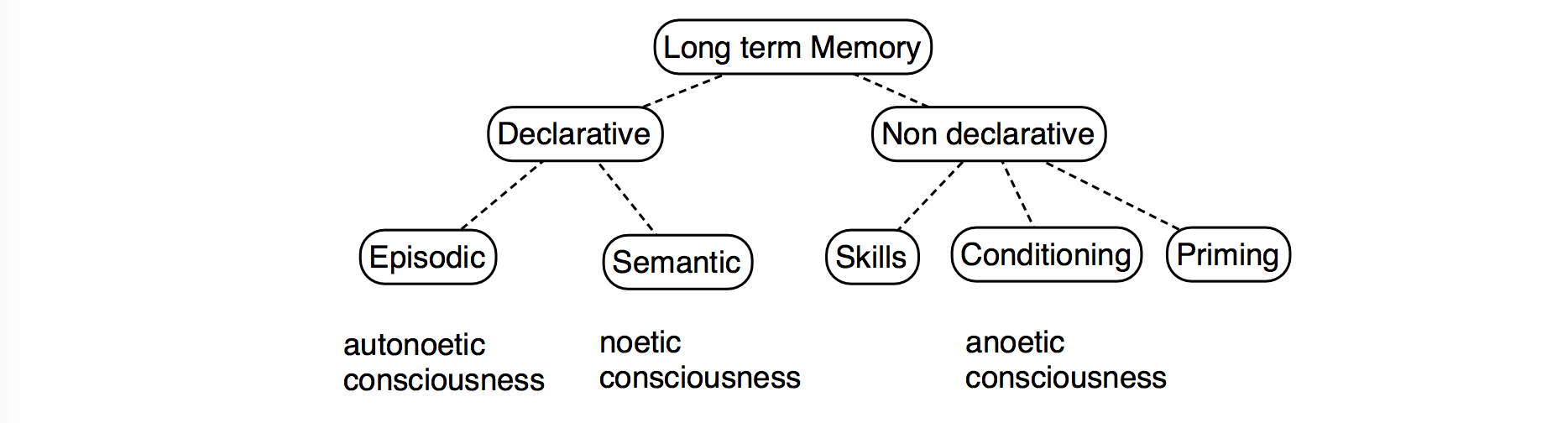}
\caption{\color{Gray} \textbf{A taxonomy that includes declarative and non declarative memory systems, including the different kinds of awareness of the subject. Although non declarative memory was not yet fully understood, compelling evidence pointed at distinct specialized components of memory outside of the episodic and semantic domains.}
}
\label{fig0} 
\end{figure}

\subsection{Non declarative memory}
There were early suggestions in the animal literature that more than just motor skills were intact after lesions of hippocampus or related structures \cite{Gaffan1974RecognitionIA, Hirsh1974TheHA,Okeefe1978TheHA}. However, these proposals greatly differed from each other, and they came at a time when the findings in experimental animals was not in agreement with findings from human amnesia and memory \cite{Milner1963EffectsOD,Milner1968FurtherAO}.

Evidences of the extent of non declarative systems in humans came from the task of reading mirror-reversed words \cite{Cohen1980PreservedLA}, a perceptual skill which amnesic patients acquired at a normal rate despite poor memory for the words that they read. Thus, perceptual skills, not just motor skills, were intact. Later, amnesiac patients with the ability to resolve stereoscopic images \cite{Benzing1989PreservedLA}, learning of an artificial grammar \cite{Knowlton1992IntactAG}, and learning of new categories of items \cite{Knowlton1993TheLO}. The accumulating data led to further arguments about the extent and nature of what is not included in the notion of declarative memory \cite{Milner1968FurtherAO,Warrington1968NewMO}.

The notion of procedural memory was originally devised to contrast with declarative memory \cite{Winograd1975FRAMERA, Cohen1980PreservedLA}. Whereas this notion appropriately describes a wide variety of skill-based kinds of learning, certain memory phenomena were clearly not declarative, but also not well accommodated by the notion of procedural memory. Subsequently, the concept of non declarative memory emerged as an umbrella which would cover human memory capacities that support skill and habit learning, perceptual priming, and other forms of behavior, which are expressed through performance rather than recollection \cite{Squire1988MemoryBS}. For example procedural skills, such as knowing how to ride a bicycle, are expressed through performance in riding a bicycle; priming memory, which underpins our ability to recognize words and other perceptual skills, is accessed implicitly as it is performed.
Non declarative memory is neither true nor false, does not require conscious reflection on the past, or event knowledge that memories are being formed by past events. Non declarative memory is shaped by experience, but unlike declarative memory which is flexible and can guide behavior in multiple contexts, the acquired knowledge in non declarative memory is thought to be rigidly organized \cite{Squire1992DeclarativeAN}.

Authors have suggested that non declarative memory is phylogenetically old. and that it may have developed as a set of special purpose learning abilities. Memory formation is then thought as the result of cumulative changes stored within the particular neural systems engaged during learning \cite{Squire2004MemorySO}. 
Most non declarative memory functions are present at the onset of human life, few age-related improvements have been found from 3 months on, and evidences have highlighted that these systems remain stable since the 9th month of life \cite{Alibali1999TheDP,Kolling2016CrossCulturalPO,Vhringer2018TheDO}.
The idea that non declarative memory remains stable as age progresses has come to be widely accepted. For instance, several studies of priming in normal aging have not reported significant decline, while cross-sectional studies employing a range of different tasks have reported non-significantly different priming between groups of young and older adults \cite{Fleischman2007RepetitionPI,Spaan2011PrimingEF}. 

Yet not all authors agree on the idea of preserved non declarative functions in older individuals, and some have argued that these inconclusive results need to be further substantiated \cite{Ward2013AgeEO}.

 \subsection{Non declarative systems}
 
 Neuroimaging studies have led researchers to attempt establishing the main neural structures identified in relation to current non declarative systems. For example, the striatum has been linked to memory of skills and habits, and active sections of the neocortex have been related to priming. Classical conditioning of skeletal musculature has an essential dependence on the cerebellum, while emotional conditioning has been correlated to activation of the amygdala \cite{Squire2004MemorySO,Squire2009MemoryAB}. 
 
 It is debated to this day the role of the hippocampus and related structures in non declarative learning. One proposal is that, whereas conscious recollection depends on the hippocampus, it is also important for unconscious memory under some circumstances \cite{Hannula2012TheHR,Shohamy2013MechanismsFW}. Other work, based on functional magnetic resonance imaging (fMRI), has implicated MTL structures in the unaware learning of sequences and other tasks with complex contingencies \cite{Chun1999MemoryDF,Schendan2003AnFS}. Considering that fMRI data cannot establish a necessary role for a particular structure, this remains a debated topic.
 
 The role of non declarative memory system in supporting syntactic processing is also still unclear. In studies with patients with Korsakoff's syndrome who display deficits in all subdomains of declarative memory, yet their non declarative memory remains intact, patients showed robust syntactic priming effects, the phenomenon in which participants adopt the linguistic behaviour of their partner \cite{Heyselaar2017TheRO}.
 
 \subsubsection{Conditioning: Skeletal responses}
 Study of nondeclarative memory began with motor skills and perceptual skills. The best understood example of nondeclarative memory in vertebrates is classical conditioning of the eyeblink response, specifically delay eyeblink conditioning. In delay conditioning, a neutral conditioned stimulus (CS), such as a tone, is presented just before an unconditioned stimulus (US), such as an airpuff to the eye. 
 The essential memory trace for the conditioned eyeblink response and other discrete conditioned motor responses was discovered is formed and stored in the cerebellar interpositus nucleus \cite{McCormick1981TheEF}.
 Critically, delay eyeblink conditioning is intact in amnesia and is acquired independently of awareness \cite{Clark1998ClassicalCA,Schacter1998MemoryAA}. Largely on the basis of work with rabbits, delay eyeblink conditioning proved to depend on the cerebellum and associated brain stem circuitry. Forebrain structures are not necessary for acquisition or retention of classically conditioned eyeblink responses. These discoveries are on of the most successful examples of localizing a memory trace within the vertebrate brain \cite{Thompson2009TheRO}. 

 \subsubsection{Priming and perceptual learning}
 The next of these to come under study was the phenomenon of priming. Priming was defined as the improvement in the ability to to detect or classify a stimulus as the result of a recent encounter with the same or a related stimulus \cite{Tulving1990PrimingAH,Schacter1998PrimingAT}. 
 
 Evidence for the distinct priming phenomenon came from both normal subjects and amnesic patients that perform normally when tests are structured using non memory kind of instructions \cite{Schacter1998PrimingAT,Tulving1990PrimingAH,Tulving1982PrimingEI}. For example, amnesic patients often performed well when they were given three-letter word stems as cues for previously presented words, event without recognizing the stimuli as having been presented before.  With conventional memory instructions (use each cue to help in remembering a recently presented word), healthy subjects outperformed the patients \cite{Graf1984ACTIVATIONMW}. Thus, it became evident that priming is an unconscious memory phenomenon and is entirely independent of the MTL.

 Priming effects are distinct from declarative memory in two other important aspects. The information acquired by priming is fully accessible only through the same sensory modality in which material was presented initially. And priming effects are short lived, since in both normal control and amnesic patients it declined after 2 hours \cite{Schacter1998PrimingAT}.

 Priming is presumably adaptive because animals evolved in a world where stimuli that are encountered once are likely to be encountered again. Perceptual priming would result advantageous as improves the speed and  fluency by which organisms interact with familiar stimuli. For example, in the case of visual priming, the posterior
visual cortex becomes more efficient at processing precisely those stimuli that have been processed recently. This plasticity occurs well before information reaches the limbic structures important for declarative memory \cite{Schacter2007RememberingTP}.
 
Evidence of independence of the phenomena of priming and the type of memory impaired in amnesia was reproduced by the studies of perceptual priming \cite{Hamann1997IntactPM,Stark2000RecognitionMA}, and conceptual priming \cite{Levy2004IntactCP}. These works showed that severely amnesic patients can exhibit fully intact priming while performing at chance on conventional recognition memory tests for the same test items.
 
 \subsubsection{Conditining: Emotional responses}
 
 Emotional conditioning is understood as system of non declarative memory that is responsible for the assessment whether an encountered stimulus has positive or negative value. In humans, associative fear learning proceeded normally after hippocampal lesions, even though the CS – US pairings could not be reported \cite{Bechara1995DoubleDO}. It has been shown that the amygdala has a critical role in fear learning, and its function and connectivity appears to be conserved widely across species. In human neuroimaging studies, the amygdala was activated not only by fear but by strongly positive emotions as well \cite{Hamann2002PositiveAN}.

The biological study of fear learning and its reversal, for instance fear extinction has considerable relevance for clinical disorders such as phobias, post-traumatic stress disorder, and other anxiety disorders \cite{Quirk2008NeuralMO}.

In addition to its importance for emotional learning, the amygdala also exerts an important modulatory influence on both declarative and non declarative memory \cite{Mcgaugh2009EmotionalHA,cahil97}. Thus, activity in the amygdala, and the effect of this activity on other structures, is responsible for the fact that emotionally arousing events are typically remembered better than emotionally neutral events.
 The importance of the amygdala for modulating memory has also been demonstrated with neuroimaging. Volunteers rated the arousing effects of either neutral scenes or emotionally distressing scenes and then took a memory test for the scenes two weeks later \cite{Cahill2004SexrelatedHL}. Increased activity in the amygdala at the time of learning was associated with higher arousal ratings for the scenes and improved accuracy on the later memory test. Interestingly, this effect occurred in the left amygdala for women and in the right amygdala for men \cite{rozendaal2011}.

 \subsubsection{Procedural: Skills and habits }

One important discovery in the 1980s was that the gradual trial-and-error learning that leads to the formation of habits was supported by the striatum \cite{Mishkin1984MemoriesAH, Packard1989DifferentialEO, Knowlton1996ANH}. Habit memory is characterized by automatized, repetitive behavior and, unlike declarative memory, is insensitive to changes in reward value \cite{Dickinson1985ActionsAH}. Tasks that assess habit learning are often structured so that explicit memorization is not useful, and individuals must depend more on intuition. In this learning, what is presumably acquired is a set of dispositions to perform a task in a particular way, and not factual knowledge about the world. Unlike declarative memory, which is flexible and can guide behavior in different contexts, the acquired knowledge in this case is rigidly organized. Habit memory subsequently became an important focus of study \cite{Liljeholm2012ContributionsOT}.
 
The difference between declarative memory and habit memory was shown for memory-impaired patients with hippocampal lesions and patients with Parkinson's disease \cite{Knowlton1996ANH}. Amnesic patients can acquire a variety of skills at an entirely normal rate. These include motor skills \cite{Brooks1976WhatCA}, perceptual and motor skills \cite{Nissen1987AttentionalRO,Cohen1980PreservedLA}, and cognitive skills \cite{Squire1990CognitiveSL}.

 Currently is known, striatal neuronal plasticity enables basal ganglia circuits to interact with other structures and thereby contribute to the processing of procedural memory \cite{Haber2000StriatonigrostriatalPI}. The cerebellum is involved in the execution of movements and the perfection of motor agility needed procedural skills. Damage to this area can impede one from relearning motor skills and recent studies have linked it to the process of automating unconscious skills during the learning phase \cite{Kreitzer2009PhysiologyAP}. The limbic system shares anatomical structures with a component of the striatum, which assumes primary responsibility for the control of procedural memory. Reward-based learning of this kind depends on dopamine neurons in the midbrain, which project to the striatum and signal the information value of the reward. The dorsolateral striatum is crucial for the development of habits in coordination with other brain regions while together with the infralimbic cortex appear to work together to support a fully formed habit \cite{Schultz2013UpdatingDR}.

\subsubsection{Non associative memory}

Non associative memory is understood as a robust form of behavioral memory, based on newly learned behavior through repeated exposure to an isolated stimulus. This form of memory is developed even in invertebrate animals who do not have a capacity for declarative memory.  It differs from associative learning in that it does not require the temporal pairing between two different sensory stimuli or between a sensory stimulus and corresponding response feedback \cite{Groves1970HabituationAD,Prescott1998InteractionsBD}. 
There are two well known types of non-associative learning: habituation and sensitization. Habituation is a decrease in response to a benign stimulus when the stimulus is presented repeatedly, while sensitization describes an augmentation of the response. Habituation is thought to be related to a decrease in the efficiency of synaptic transmission, which may be caused by a conductivity change in the membrane of the stimulated neuron. In turn, the process of sensitization may be due to a provision in transmission, whether it may be presynaptic or postsynaptic \cite{Glanzman2010CommonMO}. 

A well understood and simple form of non associative memory is olfactory habituation, in which responsiveness to stable but behaviorally non significant stimuli is decreased in relation to exposition \cite{Pellegrino2019ElectrophysiologicalIO,Wilson2013LateralEC}.

\section{Multiple memory systems}
In biology, a system is defined in terms of both structure and function. In its application to the domain of memory, multiple memory systems is understood as the set of different neurocognitive structures whose physiological workings produced the introspectively apprehensible and objectively identifiable consequences of learning and memory. The various memory systems can be distinguished in terms of its brain mechanisms, the different kinds of information they process and the principles by which they operate \cite{1994book}.

As it has been noted, early interpretations of memory assumed a unitary model that could be used in many different ways \cite{Humphreys1989DifferentWT,Roediger1989ExplainingDB}. However, in most of the history of the study of memory, dichotomous classifications of memory had been put to use, such as procedural versus declarative \cite{Cohen1984PreservedLC}, semantic and episodic \cite{Tulving1972EpisodicAS,Tulving1983ElementsOE}, habit and memory \cite{Hirsh1974TheHA,Mishkin1984MemoriesAH, Mishkin1984MemoriesAH2}, dispositional and representational \cite{Thomas1984DeficitsFR}, taxon versus locale \cite{Okeefe1978TheHA}. Although a minority, distinctions among three and even more memory systems have also been put forward \cite{Tulving1985HowMM}.

One of the earliest references to memory systems appeared in a 1972 paper by Tulving \cite{Tulving1972EpisodicAS}. There were other less known efforts in favor of memory systems such as the 1979 article by Warrington in which she discussed neuropsychological evidence supporting a distinction between STM and LTM systems, and between two kinds of LTM systems, namely event memory and semantic memory \cite{Warrington1979NeuropsychologicalEF, Tulving1985MemoryAC,Polster1991CognitiveNA}. 

In the early 1980s, the cerebellum was discovered to be essential for delay eyeblink conditioning \cite{McCormick1982InitialLO} a form of learning preserved both in animals with hippocampal lesions \cite{Schmaltz1972AcquisitionAE} and in severely amnesic patients \cite{Clark1998ClassicalCA}. Then, the striatum was identified as important for the sort of gradual, feedback-guided learning that results in habit memory \cite{Mishkin1984MemoriesAH, Packard1989DifferentialEO}. A similar contrast between declarative memory and habit memory was later demonstrated for amnesic patients and patients with Parkinson's disease \cite{Knowlton1996ANH}. Finally, it was shown that still other types of learning, which involve the attachment of positive or negative valence to a stimulus, as in fear conditioning or conditioned place preference, have an essential dependence on the amygdala \cite{Davis1992TheRO,Phillips1992DifferentialCO,Phelps2005ContributionsOT}. 

Based on evidence from human studies, these and other models attempted to embrace the large amount that has been learned about neuroanatomy, the molecular and cellular biology of synaptic change, and the organization of brain systems. Work with experimental animals, namely rats and monkeys without which a systematic study of memory would have proved impossible, also supported the notion of multiple memory systems.  

Given this wide variety of evidence related to memory phenomena and its supporting neural sustrates, in the mid 1980s the perspective abandoned an account of human memory based on a two-part dichotomy and shifted to a more complex taxonomy \cite{Tulving1985MemoryAC}. Critically, the notion of non declarative memory was introduced and conceptualized as an umbrella term referring to several memory systems supporting the wide panorama of memory related phenomena found in human and animal research \cite{Squire1988MemoryBS, 1994book}.

\begin{figure}[ht] 
\includegraphics[width=\textwidth]{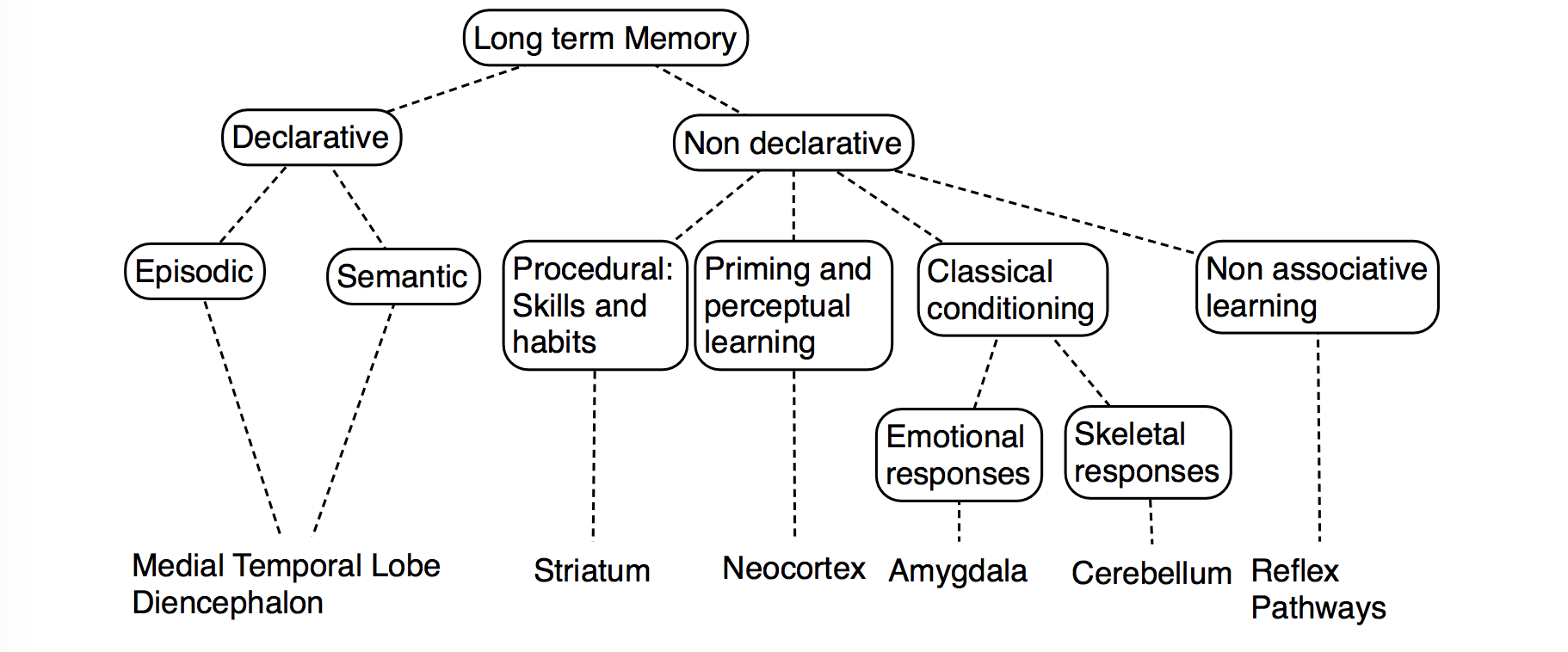}
\caption{\color{Gray} \textbf{ Taxonomy of mammalian long term memory systems according to Squire \cite{Squire2004MemorySO}. The taxonomy lists the brain structures thought to be especially important for each form of declarative and non declarative memory. 
}
}
\label{figtax} 
\end{figure}

Sherry and Schacter approached the concept of memory systems from an evolutionary perspective, proposing that different systems evolve as special adaptations of information storage and retrieval for specific and functionally incompatible purposes \cite{Sherry1987TheEO}. In their evolutionary view, Sherry and Shacter elaborate the idea of functional incompatibility, examining components of this evolution such as natural selection, heritable variation in memory, memory and reproductive success, and adaptive specialization \cite{Sherry1992SpatialMA}. The notion assumes that in order to solve different environmental problems, the animal must make feasible different requirements of the information processing strategies, and that these different requirements can call for distinct neural circuitries \cite{Schacter1996SearchingFM,Schacter2012TheFO}.

The memory systems of the mammalian brain operate independently and in parallel to support different behaviors. In some circumstances, memory systems are described as working cooperatively to optimize behavior and in other circumstances are described as working competitively. How they compete or substitute one for the other is a topic of current interest \cite{Packard2013FactorsTI,Lungu2014StriatalAH,Goodroe2018TheCN}. Packard demonstrated the parallel works of multiple memory systems in rats \cite{Packard1996InactivationOH}.% in 96 that 

When the stratial caudate nucleus was inactive, the parallel memory system supported by the hippocampus was unmasked. A similar stance has been described in humans performing a virtual navigation task that could be solved by either a spatial or nonspatial strategy \cite{Iaria2003CognitiveSD}. As training progressed participants tended to shift to a nonspatial strategy, then showing increased activity in the stratial caudate nucleus, which emerged as training progressed. 
Although many tasks can be acquired by more than one memory system, other tasks strongly favor one system over another. Prefrontal cortex may  be important in determining which memory system gains control over behavior \cite{mcdonald_hong_2013}. Several other factors increase the tendency to adopt one memory system over another, in this case, a striatal strategy. In this sense, relevant evidence has been obtained from studies of stress \cite{Schwabe2013StressAM}, and aging \cite{Konishi2013SpatialNS}. Although many tasks can be acquired by more than one memory system, other tasks strongly favor one system over another.

Similarly, a common feature of skill learning in humans is that trying to memorize, and use declarative memory, can disrupt performance, as revealed by fMRI activity in the MTL early during learning \cite{Poldrack2001CharacterizingTN}. When learning progressed, activity decreased in the MTL, and activity increased in the striatum. Moreover, when the task was modified so as to encourage the use of declarative memory, less activity was observed in the striatum and more activity was observed in the MTL.

Using these notions of multiple memory systems, researchers were increasingly more able to place theoretical speculation within a neurobiological framework, thus reaching a more accurate understanding and classification of memory.

\section*{Conclusion}

The nature of memory and its organization has been central to discussion on memory for several centuries. The evolution of the idea that there are multiple forms of memory, each supported by a distinct brain system, began in the mid 20th century and is now widely accepted and fundamental to the contemporary study of learning and memory \cite{Eichenbaum2001FromCT,Squire2004TheMT,Schacter2000MemorySO}.

The multiple memory systems framework is supported by an enormous amount of data learned in the past decades about neuroanatomy and cognitive neuroscience, and it has been able to accommodate a large variety of empirical observations of memory performance. 
It seems justifiable to conceptualize that the various memory components of memory have distinct neural architectures and operational principles, and serve different domain-specific purposes. Despite these memory systems operate independently and in parallel to support different behaviors, in some circumstances, memory systems may work cooperatively to optimize behavior and in other circumstances work competitively. It is still debated, however, how to best characterize their mutual relations and interactions. It remains beyond the scope of this article to evaluate the strengths and weaknesses of all proposed distinctions between memory systems.

A relevant implication of the multiple memory systems framework is that the therapeutic targets for various kinds of memory disorders would demand  different interventions based on the known supporting neural architectures. 
Future investigation that combine cognitive, neuropsychological and neuroimaging approaches will further our understanding of the relations among the systems that together constitute the foundation of memory. Therefore, it is foreseeable that the notion of multiple memory systems remains dynamic, and that new specifications of memory systems emerge in the future grounded on empirical evidence.

%\clearpage

\nolinenumbers

\bibliography{library}

\bibliographystyle{unsrt}

\end{document}